\documentclass[aps,nofootinbib,prl,reprint,superscriptaddress]{revtex4-1}

\pdfoutput=1
\usepackage{graphicx}

\usepackage{multirow}

\usepackage{amsmath,amssymb,mathrsfs}

\usepackage[bookmarks=false]{hyperref} %make sure it comes last of your loaded packages
\hypersetup{pdfstartview=FitH,pdfhighlight=/O,colorlinks=false}

\newcommand{\mc}[1]{\mathcal{#1}}

\newcommand{\p}{\partial}

\newcommand{\eps}{\varepsilon}

\newcommand{\mdots}{,.\,.\,,}

\begin{document}

\title{Discreteness of the volume of space from Bohr-Sommerfeld quantization}

\author{Eugenio Bianchi} 
\affiliation{Centre de Physique Th\'eorique de Luminy, case 907, F-13288 Marseille, France}
%\email{bianchi@cpt.univ-mrs.fr}
%\footnote{\scriptsize Unit\'e mixte de recherche (UMR 6207) du CNRS et des Universit\'es de Provence (Aix-Marseille I), de la M\'editerran\'ee (Aix-Marseille II) et du Sud (Toulon-Var); laboratoire affili\'e \`a la FRUMAM (FR 2291).}

\author{Hal M. Haggard}
\affiliation{Department of Physics, University of California,  Berkeley, California USA}
%\email{hal@berkeley.edu}

\begin{abstract}
A major challenge for any theory of quantum gravity is to quantize general relativity while retaining some part of its geometrical character. We present new evidence for the idea that this can be achieved by directly quantizing space  itself. We compute the Bohr-Sommerfeld volume spectrum of a tetrahedron  and show that it reproduces the quantization of a grain of space found in loop gravity.
\end{abstract}

%\begin{flushleft}
%Keywords: 
%\end{flushleft}
%\begin{flushleft}
%PACS: 04.60.Pp
%\end{flushleft}

\maketitle

%\section{Introduction}
At the Planck scale, a quantum behavior of the geometry of space is expected. Loop gravity provides a specific realization of this expectation: it predicts a granularity of space with each grain having a quantum behavior \cite{Rovelli:1994ge}. In particular, the volume of a grain of space is quantized and has a discrete spectrum with a rich structure \cite{DePietri:1996pja}. 

%%The derivation of the spectrum of the volume involves the calculation of the eigenvalues of an operator defined on the Hilbert space of \emph{intertwiners} between unitary irreducible representations of the group $SU(2)$. 

In this letter, we present a new independent road to the granularity of space and the computation of the spectrum of the volume. The derivation is based solely on semiclassical arguments applied to the simplest model for a grain of space, a Euclidean tetrahedron,  and is closely related to Regge's discretization of gravity and to more recent ideas about general relativity and quantum geometry \cite{Freidel:2010aq,Bianchi:2010gc}. The spectrum is computed by applying Bohr-Sommerfeld quantization to the volume of a tetrahedron seen as an observable on phase space. The result is accurate for large quantum numbers.

%In this letter, we present a new road to the granularity of space via semiclassical arguments and again arrive at a discrete volume spectrum for the simplest grain of space, a Euclidean tetrahedron.  This spectrum is computed by applying Bohr-Sommerfeld quantization to a classical dynamical system: the Euclidean tetrahedron endowed with the dynamics on phase space generated by its volume. 

Our central question is whether this Bohr-Sommerfeld volume spectrum, and the eigenvalues of the volume operator obtained quantizing general relativity with loop methods are related. The remarkable quantitative agreement of the two volume spectra presented here supports this idea. The result is of interest as it lends further credibility to the intricate derivation of the volume spectrum in loop gravity, showing that it matches with the elementary semiclassical approach presented here.

We begin by reviewing how convex polyhedra can be treated as dynamical systems. Then we discuss the Bohr-Sommerfeld quantization of the volume of a tetrahedron and conclude comparing our results to those found in loop gravity. \\

%\section{The phase space of polyhedra}
Two elegant mathematical results are key in what follows: Consider a convex polyhedron in three-dimensional Euclidean space. The first result is a theorem of Minkowski's that states that the areas $A_l$ and the unit-normals $\vec{n}_l$ to the faces of the polyhedron fully characterize its shape\footnote{More precisely, given a set of $N$ positive numbers $A_l$, and $N$ unit-vectors $\vec{n}_l$ satisfying the condition $\sum_l A_l \vec{n}_l=0$, there always exists a convex polyhedron having these data as areas and normals to its faces. Moreover, up to rotations $SO(3)$, the polyhedron is unique.}, \cite{Minkowski}.
We define the vectors $\vec{A}_l=A_l\,\vec{n}_l$ and call $\mc{P}_N$ the \emph{space of shapes of polyhedra} with $N$ faces of given areas $A_l$,
\begin{equation*}
\textstyle \mc{P}_N=\big\{\vec{A}_l,\,l=1\,.\,.\,N\,|\;\sum_l \vec{A}_l\,=0\,,\,\|\vec{A}_l\|=A_l\big\}/SO(3)\;.
%\label{eq:space of shapes}
\end{equation*}
The second is a result of Kapovich and Millson's that states that the set $\mc{P}_N$ has naturally the structure of a \emph{phase space}, \cite{Kapovich}. The Poisson brackets between two functions $f(\vec{A}_l)$ and $g(\vec{A}_l)$ on $\mc{P}_N$ are 
\begin{equation}
\big\{f,g\big\}=\sum_l\, \vec{A}_l\cdot \big(\,\frac{\p f}{\p \vec{A}_l}\times \frac{\p g}{\p \vec{A}_l}\,\big)\;.
\label{eq:PB}
\end{equation}
These brackets arise (via symplectic reduction) from the rotationally-invariant Poisson brackets between functions $f(\vec{A}_l)$ on $(S^2)^N$. Thus we have that convex polyhedra with $N$ faces of given areas form a $2(N-3)$ dimensional phase space \cite{Bianchi:2010gc}.

Canonical variables on this phase space can be chosen as follows: consider the set of vectors $\vec{p}_k=\sum_{l=1}^{k+1} \vec{A}_l$, where $k=1\mdots N-3$; we define the coordinate $q_k$ as the angle between the vectors $\vec{p}_k\times \vec{A}_{k+1}$ and $\vec{p}_k\times \vec{A}_{k+2}$, and the momentum variable $p_k=\|\vec{p}_k\|$ as the norm of the vector $\vec{p}_k$. From (\ref{eq:PB}), it follows that these are canonically conjugate variables, $\{q_k,p_{k'}\}=\,\delta_{k k'}\;$.\\

\begin{figure}[t]
\includegraphics[height=155pt]{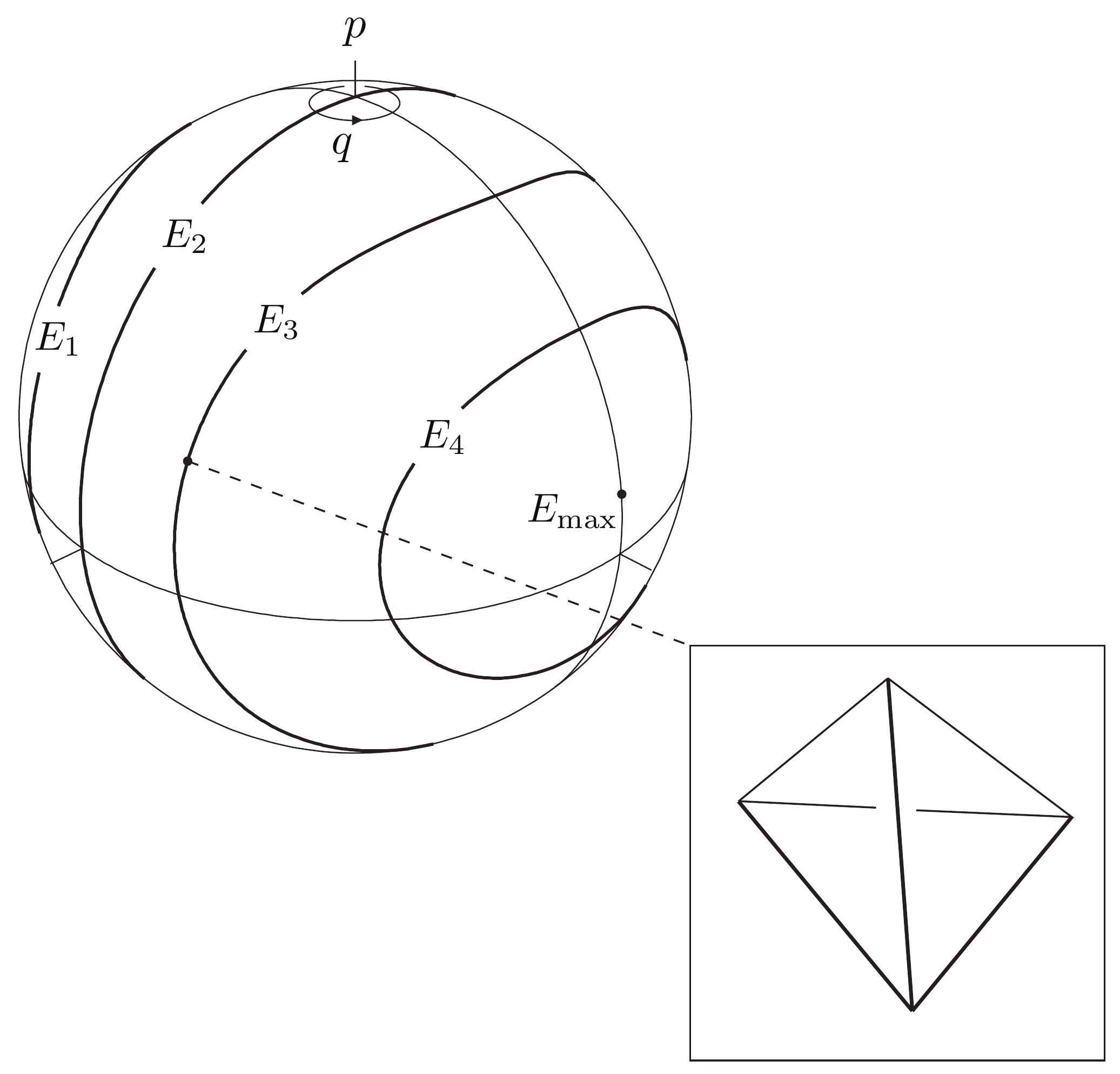}
\caption{The phase space $\mc{P}_4$ of a tetrahedron. The position $q$ corresponds to the longitude on the sphere, the momentum $p$ to the latitude. Some orbits of energy $E_n$ are shown. The volume of the tetrahedron is constant along the orbits. }
\label{fig:orbits}
\end{figure}

In the simplest non-trivial case, $N=4$, the phase space is two-dimensional, has the topology of a sphere $S^2$, and describes the shape of a tetrahedron with faces of given area, Fig. \ref{fig:orbits}. The coordinate $q$ measures the angle between two opposite edges of the tetrahedron. The conjugate momentum $p=\|\vec{A}_1+\vec{A}_2\|$ measures the dihedral angle between two faces of the tetrahedron. It varies in the interval $[p_{\text{min}},p_{\text{max}}]$, with $p_{\text{min}}= \text{max}(|A_1-A_2|,|A_3-A_4|)$ and $p_{\text{max}}=\text{min}(A_1+A_2,A_3+A_4)$, \cite{Barbieri:1997ks}.

The volume $V$ of the tetrahedron is a function on this phase space, $\mc{P}_4$, and is given by
\begin{equation}
V=\frac{\sqrt{2}}{3}\sqrt{|H(q,p)|}\;,
\label{eq:V=sqrt H}
\end{equation}
where $H(q,p)=\vec{A}_1\cdot (\vec{A}_2 \times \vec{A}_3)$ is the triple product of the normals to its faces.\\
%% short version %% \footnote{Because of the closure condition $\sum_l \vec{A}_l=0$ and up to a sign, the function $H$ is independent of the choice of three out of the four faces of the tetrahedron.}\\

We derive the spectrum of the volume under the following two physical assumptions:
(i) the first is that, in a quantum theory of gravity, the full dynamics induces on a grain of space -- a tetrahedron -- the natural rotationally-invariant Poisson brackets (\ref{eq:PB}) discussed above;
(ii) the second assumption is that Bohr-Sommerfeld quantization can be applied to the volume observable V on the phase space $\mc{P}_4$. 

In particular, (ii) restricts the possible values of the area of the faces of the tetrahedron, \cite{Aquil:3j2007}. We assume that they are of the form
\begin{equation}
A_l=(j_l+\frac{1}{2})\, \hbar\;,
\label{eq:area}
\end{equation}
where $j_l$ is a half-integer, $j_l=\frac{1}{2},1,\frac{3}{2},\ldots$, and $\hbar$ is Planck's constant.\footnote{We work in units $c=1$ and Newton's constant $G=1$, but $\hbar$ will be kept explicit. As a result, the areas $A_l$ have the same dimensions as Planck's constant $\hbar$. We also set $8\pi \gamma =1$, where $\gamma$ is the Immirzi parameter proper to loop gravity.} This condition guarantees that the total symplectic area of phase space 
\begin{equation}
\int_{\mc{P}_4} dq\wedge dp =  2\pi\,(p_{\text{max}}-p_{\text{min}}) \; \equiv \;2\pi \hbar\,d\;
\end{equation}
is an integer multiple $d$ of $2\pi\hbar$. Moreover, it is consistent with the semiclassical limit of the area spectrum in loop gravity.\footnote{In loop gravity, the area spectrum is $A_l=\sqrt{j_l(j_l+1)}\,\hbar$. For large $j_l$, the area spectrum coincides with (\ref{eq:area}).} We call $j_l$ the \emph{spin} of the face $A_l$.

In applying Bohr-Sommerfeld quantization, it is convenient to regard the function $H(q,p)$ as the Hamiltonian of the system. Expressed in terms of the canonical variables $q$ and $p$, it is given by
\begin{align*}
H(q,p)&=\;\frac{1}{4p}\sqrt{[p^2-(A_1-A_2)^2][p^2-(A_3-A_4)^2]}\;\times\\
&\times\;\sqrt{[p^2-(A_1+A_2)^2][p^2-(A_3+A_4)^2]}\;\sin q\; .
\label{eq:H(q,p)}
\end{align*}
The evolution of the shape of the tetrahedron is described by Hamilton's equations, $\dot{q}=\frac{\p H}{\p p}\,$, $\,\dot{p}=-\frac{\p H}{\p q}\,$. They describe closed orbits of constant energy $H(q,p)=E$ and period $T(E)$ (see Fig. \ref{fig:orbits}). The volume $V$ of the tetrahedron is constant along these orbits.\\

%\section*{Bohr-Sommerfeld quantization}
We briefly recall Bohr-Sommerfeld quantization and show how it can be applied to our system to determine the spectrum of the volume.

\begin{figure}[t]
   \includegraphics[height=155pt]{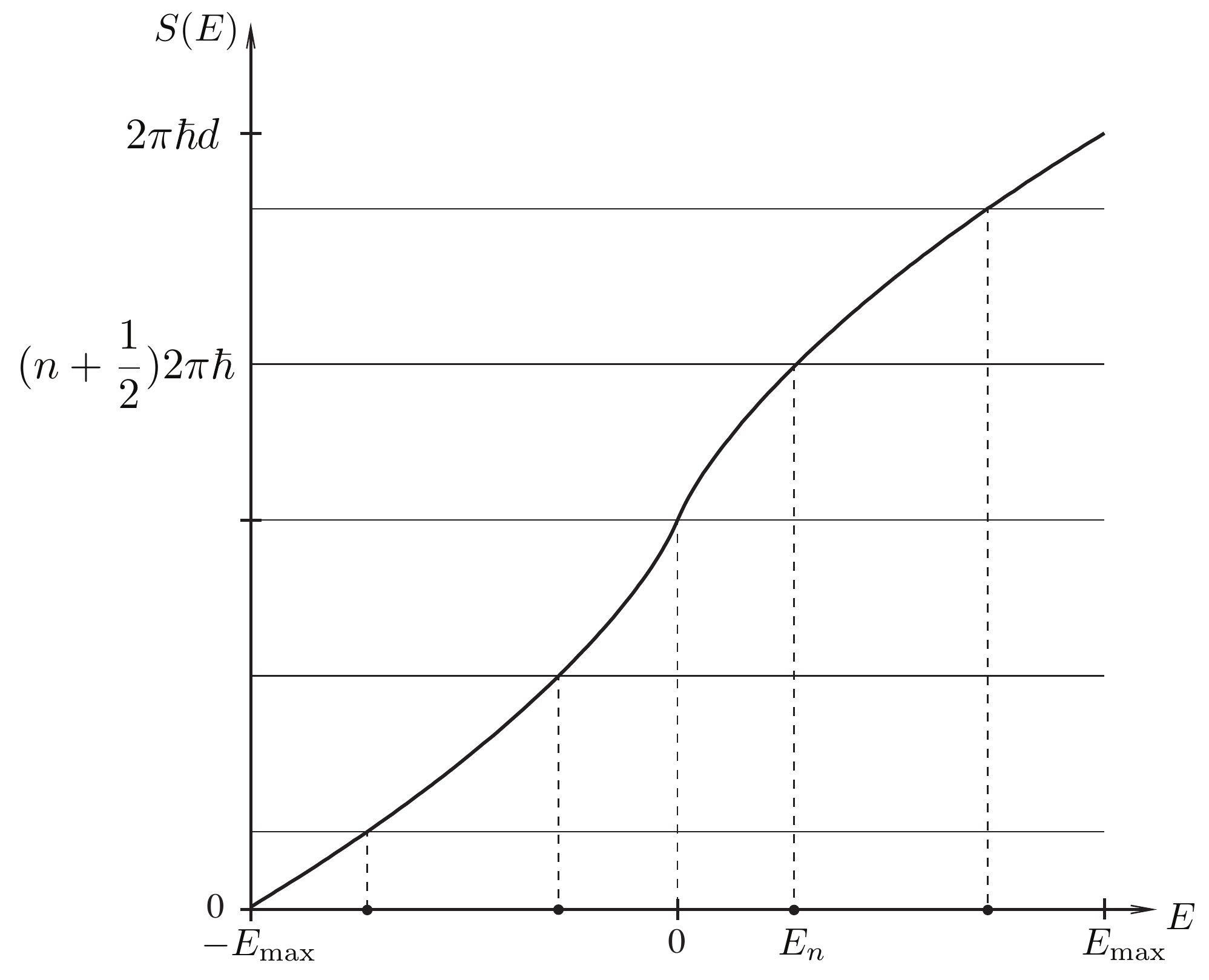} 
   \caption{Plot of the Jacobi action integral $S(E)$. The energy-levels $E_n$ shown satisfy the Bohr-Sommerfeld quantization condition. The corresponding orbits are shown in Fig. \ref{fig:orbits}.}
   \label{fig:action}
\end{figure}

\begin{figure*}[t]
   \includegraphics[height=155pt]{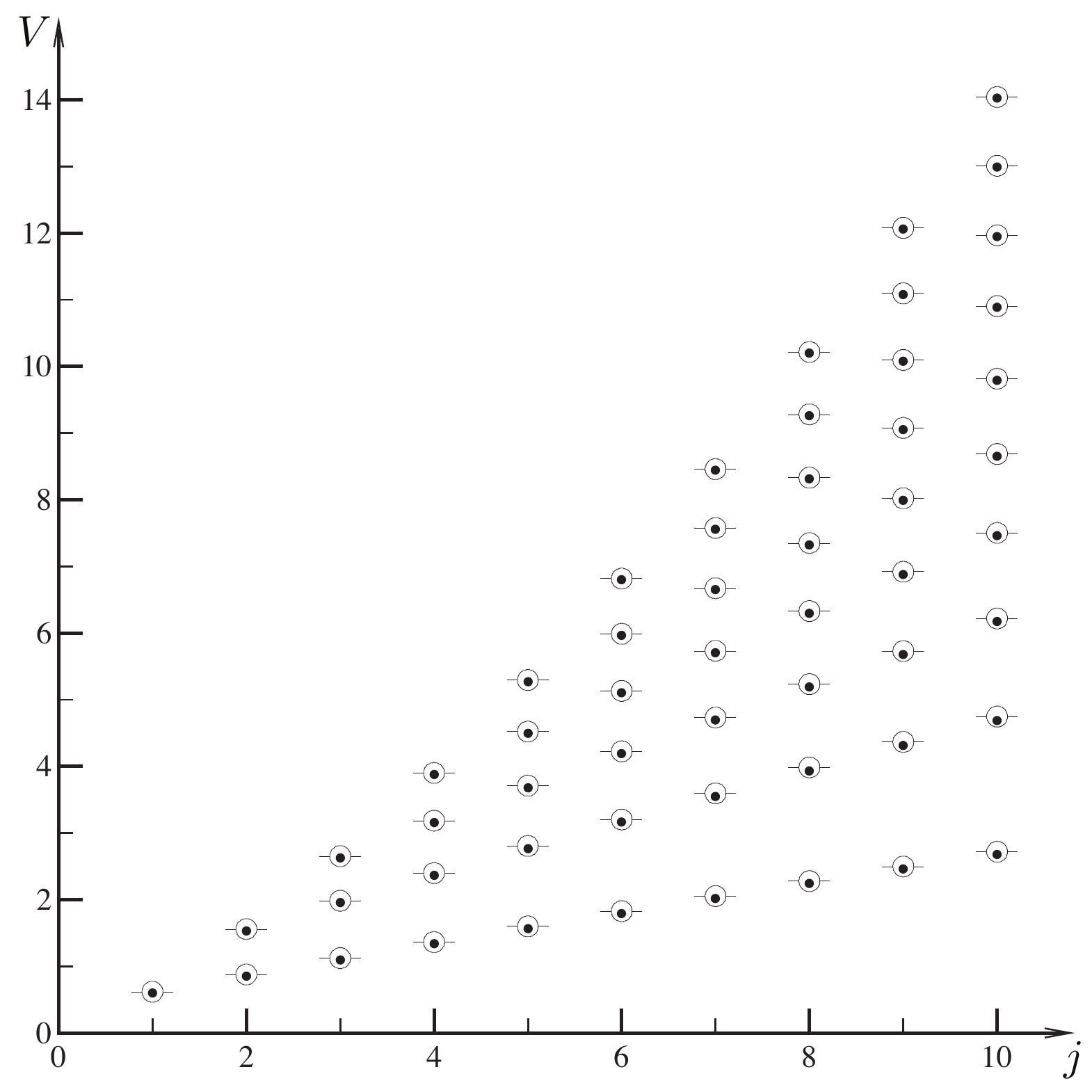} \hspace{80pt}
   \includegraphics[height=155pt]{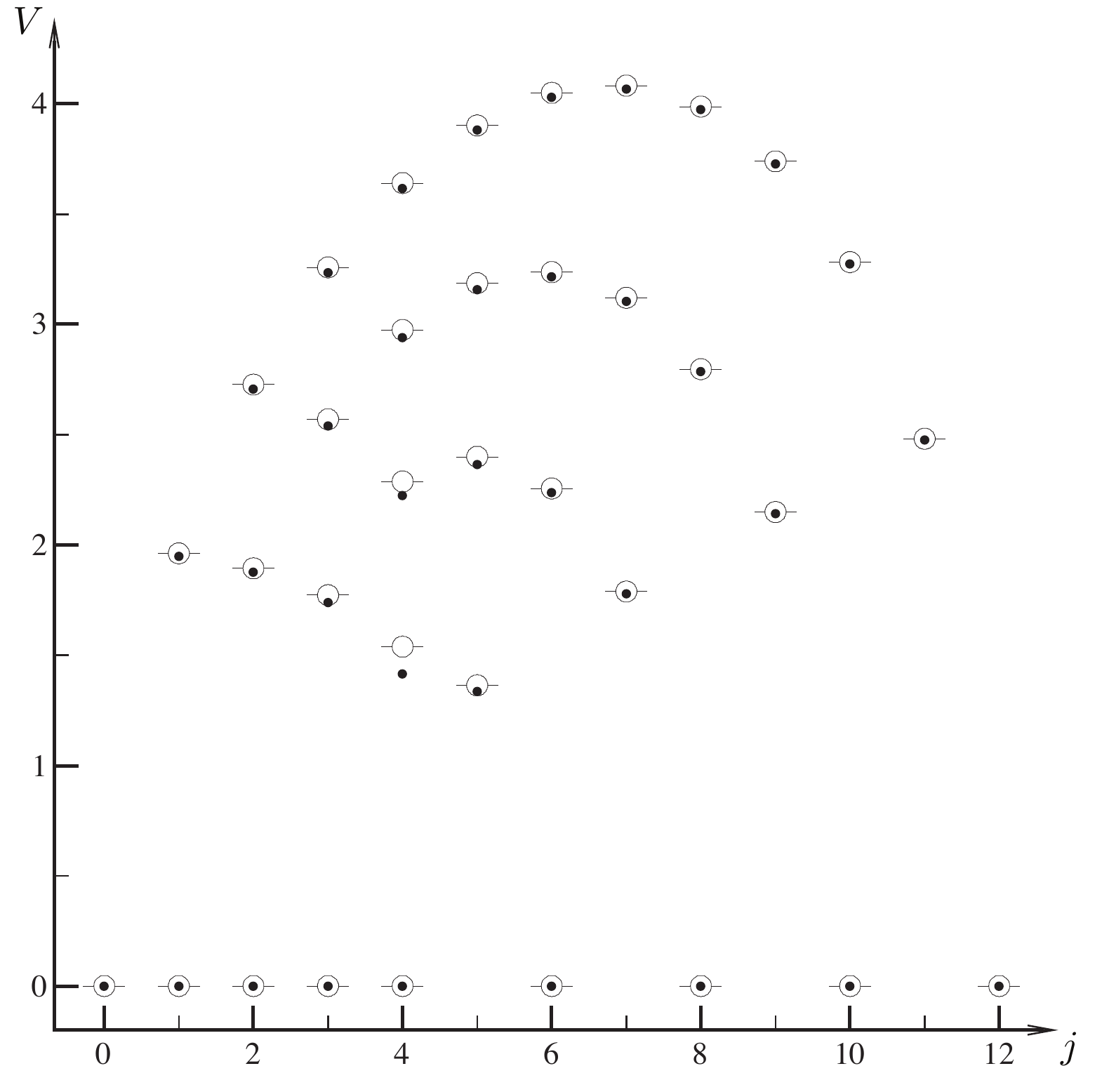} 
   \caption{Volume spectrum. On the left: configuration with spins $\{j,\,j,\,j,\,j+1\}$. On the right: configuration with spins $\{4,\,4,\,4,\,j\}$ and $j$ varying in its allowed range. The Bohr-Sommerfeld values of the volume of a tetrahedron are represented as \emph{dots}, the eigenvalues of the loop-gravity volume operator as \emph{circles}. Notice the quality of the matching of the two.}
    \label{fig:PIII}
\end{figure*}

Before the birth of quantum mechanics, Planck hypothesized that the phase space of a system is \emph{quantized} in cells of area $2\pi\hbar$. The argument was made more precise by Bohr and Sommerfeld: the only allowed orbits for a periodic system are the ones that encircle an area in phase space that is a multiple of Planck's constant. The quantity that measures the area as a function of the energy $E$ of the orbit is the Jacobi action integral $S(E)$,
\begin{equation}
S(E)=\int_{H(q,p)\leq E}dq\wedge dp\;=\oint_E p\,dq\;.
\label{eq:S(E)}
\end{equation}
For our system, this quantity can be computed and is plotted in Fig. \ref{fig:action}. The Bohr-Sommerfeld quantization condition requires that
\begin{equation}
S(E_n)=\big(n+\frac{1}{2}\big)\;2\pi \hbar\;,
\label{eq:BS}
\end{equation}
where $n$ is an integer. This condition identifies the allowed values of the energy $E_n$ for a stationary state of the system. From the modern perspective these values provide a semiclassical approximation to the eigenvalues of the Hamiltonian. The corrections can be computed in a WKB expansion \cite{Cargo}. A rough estimate is that they are negligible for $E_n \gg 2 \pi \hbar/T(E_n)$.

%Here, we apply this same quantization rule to the tetrahedral phase space $\mathcal{P}_4$ with the dynamics given by the Hamiltonian $H(q,p)$. We compute the allowed energies $E_n$ and 

The spectrum of the volume is simply given by the values 
\begin{equation}
v_n=\frac{\sqrt{2}}{3}\sqrt{|E_n|}\;,
\label{eq:vn}
\end{equation}
as follows from (\ref{eq:V=sqrt H}). We compute explicitly the levels $v_n$ and compare them to those computed in full loop gravity.\\

%\footnote{We are deriving the spectrum of the spatial volume in quantum gravity using methods that are older than quantum mechanics itself. In the `old quantum theory', the quantization condition (\ref{eq:BS}) was in fact based on the following physical assumptions. There are systems in nature whose energy spectrum was then known to be discrete and equispaced (for instance the harmonic oscillators in Planck's work on the black body spectrum, or in Einstein and Debye's work on specific heats). If we slowly modify the parameters of these systems, the energy spectrum changes but its discreteness is not detroyed. Lorentz and Einstein asked if there is a classical version of this robustness of the spectrum: is there a classical quantity that is preserved in a slow -- adiabatic -- change of the external parameters of such systems? This observation led Ehrenfest to identify \emph{adiabatic invariants} as the quantities that, on allowed orbits, are realized in multiples of $2\pi \hbar$. The Jacobi action integral $S(E)$ can be shown to be an adiabatic invariant for the system, and provides a natural candidate to extend Planck's quantization of the energy of the harmonic oscillator to other periodic systems, and in particular to the Hydrogen atom.}\\

For given a choice of spins $\{j_1,j_2,j_3,j_4\}$, and thus of areas, the Jacobi action $S(E)$ can be computed explicitly:
\begin{equation}
\label{eq:Action}
S(E) = \left( \sum_{i=1}^{4} a_i\, K(m)-\sum_{i=1}^{4} b_i\, \Pi (\alpha_i^{2}, m)\right) E\;,
\end{equation}
here $K$ and $\Pi$ are the complete elliptic integrals of the first and third kinds respectively and depend on the elliptic parameter $m$ and characteristics $\alpha_i^{2}$. These five parameters and the eight parameters \{$a_i,b_i$\} depend on the roots of an auxiliary quartic equation. The coefficients of this quartic equation are completely specified by the four spins $j_l$ and the energy $E$, thus through their dependence on the roots all of the parameters of Eq. \eqref{eq:Action} come to depend on the energy. By performing a numerical inversion of this formula we obtain the $E_n$ and the corresponding $v_n$ and are able to compare with calculations in loop gravity.
 
In Fig. \ref{fig:action} we consider a tetrahedron with faces of spin $j_l=\{2,2,2,2\}$. The Bohr-Sommerfeld condition selects $d$ allowed energy-levels $E_n$. The number of levels, $d=5$ in this example, is given by the total symplectic area of phase space. The allowed values $v_n$ of the volume of the tetrahedron are obtained via Eq.\eqref{eq:vn}. The non-vanishing values $v_n$ are twice degenerate; there are two orbits in phase space with the same value of the volume (see Fig.\ref{fig:orbits}).

%The corresponding orbits in phase space are shown in Fig. \ref{fig:orbits}. The allowed values $v_n$ of the volume of the tetrahedron are obtained via Eq.\ref{eq:vn}. The full spectrum of the Bohr-Sommerfeld volume is obtained considering all the allowed configurations of spins $j_l$. 

In summary, if we assume that, classically, space is made up of a collection of tetrahedra (as for instance in Regge's discretization of gravity), then the Bohr-Sommerfeld condition predicts that spatial volume is quantized and its spectrum can be derived.\\

%\section*{Relation to the loop gravity volume}
In loop gravity, a grain of space is represented by a node of the graph of a spin-network state. More precisely, to a node having $N$ links labeled by spins $j_l$, we associate a Hilbert space $\mc{H}_N$ known as intertwiner space. This is the space of invariants in the tensor product of $N$ representations $D^{(j_l)}$ of the group $SU(2)$,
\begin{equation}
\mc{H}_N=\text{Inv}(D^{(j_1)}\otimes \cdots \otimes D^{(j_N)})\;.
\label{eq:HN}
\end{equation}
This space is $d$-dimensional with $d$ greater than or equal to one only if $N\geq4$. The volume $V$ of a grain of space is an operator on the Hilbert space $\mc{H}_N$. This operator is obtained by regularizing and quantizing the $3$-metric $h$ in the classical expression 
\begin{equation}
V=\int_R d\vec{x}\,\sqrt{h}
\end{equation}
for the volume of a region of space $R$. In the case $N=4$, the volume operator is simply given by
\begin{equation}
V=\frac{\sqrt{2}}{3}\sqrt{|\eps_{ijk}\,J^i_1\, J^j_2\, J^k_3\,|}\;,
\label{eq:Vop}
\end{equation}
where $J^i_l$ is the generator of $SU(2)$ in the representation $D^{(j_l)}$. Once a quadruplet of spins $\{j_1,j_2,j_3,j_4\}$ and a basis in $\mc{H}_N$ have been chosen, the operator $V$ reduces to a $d\times d$ matrix. Computing the volume operator's spectrum amounts to finding the $d$ eigenvalues of this matrix, a task that can be done numerically \cite{DePietri:1996pja}.

%In Fig.\ref{fig:PIII}, we report the eigenvalues of the volume operator for the Hilbert space $\mc{H}_4$ with spins $\{j,\,j,\,j,\,j+1\}$. The dimension of the Hilbert space is $d=2j$, and the levels shown in the figure are twice degenerate. On the same figure, we report also the Bohr-Sommerfeld spectrum of the volume of a tetrahedron with the same choice of spins. The degeneracy of levels is given by the number of orbits in phase space having the same volume, i.e. $2$ orbits related by parity. The data sets agree qualitatively and quantitatively already at very low spins.

 In Figure \ref{fig:PIII}, we consider a list of spin quadruplets and compare the two volume spectra. The two data sets are in good agreement both qualitatively and quantitatively even for small spins. To better appreciate the accuracy of this agreement, we report some numerical data in the Table.

The reason for a relation between the two volume spectra can be traced back to recent developments on (twisted) discrete geometries in loop gravity \cite{Freidel:2010aq,Bianchi:2010gc}. In particular, the assumption (ii) about the Poisson brackets (\ref{eq:PB}) is the classical version of the non-commutativity of fluxes of the parallel transported electric field in loop gravity, and descends from the canonical phase space of general relativity formulated in Ashtekar's variables, \cite{Rovelli:2004tv}.\\

The Bohr-Sommerfeld approach taken here provides a new method for understanding many aspects of the rich structure of the volume spectrum in loop gravity. This is important because a deep understanding of the spectra of geometrical operators provides fertile ground for developing phenomenological tests of loop gravity. 

We briefly describe several results arising from the Bohr-Sommerfeld quantization: The value of the largest eigenvalue of the volume in $\mc{H}_4$ can be explained as the volume of the largest tetrahedron in $\mc{P}_4$. Moreover, at large quantum numbers, the levels of the volume are observed to be equispaced. This fact can be understood in terms of Bohr's correspondence principle: the spacing $\Delta V$ is given by $\frac{2\pi}{T}$, where $T$ is the period of the classical orbits at large volume. 

In loop gravity, the discrete spectra of geometrical observables provide a physical Planck-scale cut-off that renders the theory finite in the ultraviolet \cite{Rovelli:2004tv}. An important question is whether there exists a volume gap, that is a discrete gap, above zero, in the volume spectrum for all spins. We have investigated this question in $\mc{P}_4$ and find that, for a given choice of spins, i.e. of $A_l$, the lowest non-vanishing level of the Bohr-Sommerfeld volume spectrum is given by
\begin{equation}
v_{\textrm{min}} \simeq c\;\sqrt{\hbar}\; (A_1 A_2 A_3 A_4)^{1/4},
\end{equation}
where $c$ is $2/3$ for odd $d$ and $\sqrt{2}/3$  for even $d$. This result is obtained by expanding the Jacobi action around the orbits of longest period. Those phase spaces $\mc{P}_4$ containing degenerate tetrahedra require special care as there are orbits of infinite period. Nevertheless, they can be treated using the analytic expression of $S(E)$ in terms of elliptic functions.
%For instance, for equal spins $j_l=j_0$, we find that 
%\begin{equation}
%v_{\textrm{min}} \simeq  \frac{\frac{2\sqrt{\pi}}{3 \sqrt{3}} j_0}{\sqrt{\ln{\left( \frac{6 e j_0}{\pi} \right)}+\ln{\left(\ln{\left( \frac{6 e j_0}{\pi} \right)}\right)} +\cdots } },
%\end{equation}
These results will be discussed in detail in a forthcoming paper.\\

\begin{table}
\begin{ruledtabular}
\begin{tabular}{cccc}
\multicolumn{4}{c}{Table: Volume spectrum} \\[.2em]
\hline 
$j_1\ j_2\ j_3\ j_4$ & 
 Loop gravity  & Bohr-Sommerfeld & Accuracy \\
\noalign{\smallskip} \hline
\multirow{1}{*}{$\frac{1}{2}\ \frac{1}{2}\ \frac{1}{2}\ \frac{1}{2}$}&0.310& 0.252& 19\%\\
\noalign{\smallskip} \hline 
\multirow{1}{*}{$\frac{1}{2}\ \frac{1}{2}\;\, 1\;\, 1$}&0.396&0.344& 13\%\\
\noalign{\smallskip} \hline
\multirow{1}{*}{$\frac{1}{2}\ \frac{1}{2}\ \frac{3}{2}\ \frac{3}{2}$}&0.464& 0.406& 12\%\\
\noalign{\smallskip} \hline
\multirow{1}{*}{$\frac{1}{2}\ 1\ 1\ \frac{3}{2}$}&0.498&0.458& 8\%\\
\noalign{\smallskip} \hline
\multirow{2}{*}{1\ 1\ 1\ 1}&0&0&exact\\
&0.620&0.566& 9\%\\
\hline
\multirow{1}{*}{$\frac{1}{2}\ \frac{1}{2}\ 2\ 2$}&0.522&0.458& 12\%\\
\noalign{\smallskip} \hline
\multirow{1}{*}{$\frac{1}{2}\ 1\ \frac{3}{2}\ 2$}&0.577&0.535& 7\%\\
\noalign{\smallskip} \hline
\multirow{1}{*}{$1\ 1\ 1\ 2$}&0.620&0.598& 4\%\\
\hline
%\multirow{1}{*}{$\frac{1}{2}\ \frac{3}{2}\ \frac{3}{2}\ \frac{3}{2}$}&0.620&0.598& 4\%\\
%\noalign{\smallskip} 
%\hline
%\multirow{2}{*}{$1\ 1\ \frac{3}{2}\ \frac{3}{2}$}&0&0&exact\\
%&0.753&0.707& 6\%\\
%\hline
&\multicolumn{2}{c}{$\cdots$}&\\
\hline
\multirow{6}{*}{$6\ 6\ 6\ 7$}&1.828&1.795&1.8\%\\
&3.204&3.162&1.3\%\\
&4.225&4.190& 0.8\%\\
&5.133&5.105&0.5\%\\
&5.989&5.967&0.4\%\\
&6.817&6.799&0.3\%\\
\end{tabular}
\end{ruledtabular}
%\caption{Volume spectrum.}
\label{Table}
\end{table}

Bohr-Sommerfeld quantization offers a completely new perspective on the discreteness of volume in loop gravity. We have shown that it is quantitatively accurate, and that it provides an elementary account of various features of the spectrum. 

Using the semiclassical methods of \cite{Aquil:3j2007}, the eigenvectors of the volume can be computed in a WKB expansion. The same method can be applied to other geometrical operators, as well as to the alternative versions of the volume operator considered in the literature. When $N>4$, the phase space $\mc{P}_N$ has dimension greater than two. A preliminary analysis of the case $N=5$ indicates that, while the volume orbits may be chaotic, the dynamics can still be practically investigated numerically. This opens up the intriguing possibility for exploring quantum chaos in the volume spectrum of loop gravity.\\

We thank C.~Rovelli and R.~Littlejohn for useful discussions. This work was supported by a Marie Curie Fellowship (E.B.) and by a University of California, Berkeley dissertation year fellowship (H.M.H.).
\providecommand{\href}[2]{#2}\begingroup\raggedright\endgroup

\end{document}